\documentclass[twocolumn]{emulateapj}
\usepackage{graphicx}
\usepackage{dcolumn}
\usepackage{color}

\def\arcsec{\hbox{$^{\prime\prime}$}}

\begin{document}

\title{Discovery of a  population of bulgeless galaxies with extremely red Mid-IR colors: Obscured AGN activity in the Low Mass Regime?}

\author{S.~Satyapal\altaffilmark{1}, N.~J.~Secrest\altaffilmark{1}, W.~McAlpine\altaffilmark{1}, S.~L.~Ellison\altaffilmark{2}, J.~Fischer\altaffilmark{3}, J.~L.~Rosenberg\altaffilmark{1}}

\altaffiltext{1}{George Mason University, Department of Physics \& Astronomy, MS 3F3, 4400 University Drive, Fairfax, VA 22030, USA; satyapal@physics.gmu.edu}

\altaffiltext{2}{Department of Physics and Astronomy, University of Victoria, Victoria, BC V8P 1A1, Canada}

\altaffiltext{3}{Naval Research Laboratory, Remote Sensing Division, 4555 Overlook Ave SW, Washington, DC 20375, USA}

\begin{abstract}
In contrast to massive, bulge hosting galaxies, very few supermassive black holes (SMBHs) are known in either low mass, or bulgeless galaxies.  Such a population could provide clues to the origins of SMBHs and to secular pathways for their growth. Using the all-sky {\it Wide-Field Infrared Sky Explorer} ({\it WISE}\hspace{1 mm}) survey, and bulge-to-disk decompositions from the Sloan Digital Sky Survey (SDSS) Data Release 7, we report the discovery of a population of local (z$<$0.3) bulgeless disk galaxies with extremely red mid-infrared colors highly suggestive of a dominant active galactic nucleus (AGN), despite having no optical AGN signatures in their SDSS spectra. Using various mid-infrared selection criteria from the literature, there are between 30 to over 300 bulgeless galaxies with possible AGNs. Other known scenarios that can heat the dust to high temperatures do not appear to explain the observed colors of this sample. If these galaxies are confirmed to host AGNs, this study will provide a breakthrough in characterizing the properties of SMBHs in the low bulge mass regime and in understanding their relation with their host galaxies. Mid-infrared selection identifies AGNs that dominate their host galaxy's emission and therefore reveal a different AGN population than is uncovered by optical studies. We find that the fraction of all galaxies identified as candidate AGNs by WISE is highest at lower stellar masses and drops dramatically in higher mass galaxies, in striking contrast to the findings from optical studies.

\end{abstract}


\section{Introduction}
													
It is now well-established that supermassive black holes (SMBHs) $\sim{10^{6}-10^{9}}$~M$_{\sun}$ reside in the centers of bulge-dominated galaxies in the local Universe and that their black hole masses and host galaxy bulge mass are strongly correlated \citep[e.g.][]{gebhardt2001, ferrarese2000}.  These observations have formed the basis of the widely held view that galaxies and their central black holes are fundamentally linked, a link that can most simply be understood if galaxy mergers induce bulge growth and feed the central black hole in concert \citep[e.g.][]{kauffmann2003,ellison2011}. While the SMBH and host galaxy properties in the high bulge mass regime have been studied extensively, very little is known about the existence and properties of SMBHs in galaxies with low masses and/or those with small bulges. This is a significant deficiency since study of this population allows us to gain an understanding of merger-free pathways to black hole growth.  Furthermore, the occupation fraction and properties of SMBHs in galaxies with low masses and in those with no evidence for a bulge  provide one of the only observational constraints on the origin and efficiency of SMBH seeds, thought to have formed at high $z$ \citep[e.g.][]{ volonteri2010}. Since SMBHs in massive bulge-dominated galaxies have undergone significant accretion through multiple dynamical interactions over cosmic history, any information on the seed population will not be retained.  In contrast, galaxies that lack significant bulges have undergone a more secular evolution and therefore the mass distribution and occupation fraction of SMBHs in these galaxies contains clues about the original seed population, allowing us to discriminate between lower mass seeds formed from stellar remnants or massive seeds formed directly out of dense gas \citep[e.g.][]{vanwassenhove2010}. The study of black holes at the low bulge mass regime is therefore crucial to our understanding of both the origin of SMBHs and their growth and connection to galaxy evolution.

Unfortunately, finding and studying the properties of SMBHs in a significant sample of galaxies with low bulge mass is challenging since the black hole mass is expected to be small and therefore impossible to discover dynamically except in the few nearest galaxies. A significant sample of SMBHs in low mass hosts can therefore only be detected if they are accreting.  On the basis of optical spectroscopic observations, active galactic nuclei (AGNs) are found almost exclusively in massive bulge-dominated hosts, with the fraction of galaxies with optical signs of accretion dropping dramatically at stellar masses log(M/M$_{\odot}$)  $<10$~\citep[e.g.][]{kauffmann2003}.  One possible reason for this deficit of AGNs at low galaxy mass is that a putative AGN in a galaxy with a minimal bulge is likely to be both energetically weak and possibly deeply embedded in the center of a dusty late-type galaxy.  As a result, the optical emission lines may be dominated by star formation regions and/or considerably attenuated by dust.  This leaves open the possibility that there is a hidden population of accreting black holes in small-bulge systems.  To date, there exist only small samples of AGNs discovered by optical surveys in small-bulge galaxies and far fewer with no evidence for any bulge \citep[e.g.,][]{filippenko2003,barth2004,greene2007, jiang2011,simmons2013}.

Despite several X-ray and infrared studies aimed at searching for AGNs in the low bulge mass regime \citep[e.g.,][]{satyapal2007,satyapal2008,satyapal2009,gliozzi2009,mcalpine2011,secrest2012,shields2008,ghosh2008,dewangan2005,desroches2009,reines2011,araya2012}, there are still only a small number of purely bugleless galaxies with evidence for SMBHs known in the Universe.  Interestingly, several of the nearby bulgeless disk galaxies with high spatial resolution observations show evidence for prominent nuclear star clusters (NSCs) possibly suggesting an association of SMBHs and NSCs in the complete absence of a bulge \citep{satyapal2009}. In contrast, the dwarf galaxy He 2-10 shows no evidence of a NSC but shows a clear signatures of a SMBH as well as clear signs of a recent interaction \citep{reines2011}. With only a few examples, it is impossible to address important questions such as: Why do some completelyÊ bulgeless disk galaxies have AGNs and some do not?Ê Are the presence and properties of the central black hole related in any way to the properties of the host galaxies in galaxies with truly no bulge?  What is the relationship between NSCs and SMBHs?  What is the occupation fraction of SMBHs in truly bulgeless galaxies?  A statistically significant sample of low bulge and bulgeless galaxies with confirmed SMBHs is needed in order to make significant progress in the study of SMBH formation, growth and the connection with galaxy evolution.

Since AGNs in bulgeless galaxies are very rare, it is impossible to find a sizeable sample of such objects by carrying out X-ray or mid-infrared spectroscopic observations of small samples of bulgeless galaxies. The all-sky survey carried out by the {\it Wide-field Infrared Survey Explorer} ({\it WISE}\hspace{1 mm})~\citep{wright2010}~has opened up a new window in the search for optically hidden AGNs in a large number of galaxies based on their characteristically very red mid-infrared colors \citep{stern2012}. In this paper, we report the discovery of a population of nearby bulgeless galaxies discovered by  {\it WISE} to have extreme red mid-infrared colors indicative of hot dust usually thought to be heated by AGN. In Section 2, we describe the selection of our  bulgeless galaxy sample, followed by a discussion of their {\it WISE} colors in Section 3. In Section 4, we describe the properties of the subset of bulgeless galaxies in the sample identified by {\it WISE} with red mid-IR colors suggestive of AGNs. In Section 5, we discuss alternative scenarios that can produce red mid-IR colors in galaxies., followed by a discussion of supporting multiwavelength evidence for AGNs in bulgeless and low mass galaxies.  In Section 7, we explore the dependance of {\it WISE} AGN selection on stellar mass for the full range of stellar masses, followed by a summary and implications of our results  in Section 8.

\section{Bulgeless Galaxy Sample}

Our sample of bulgeless galaxies is drawn from the catalog of bulge/disk decompositions of 1.12 million galaxies from the Sloan Digital Sky Survey (SDSS) Data Release Seven from \citet{simard2011}. Here, two dimensional, point-spread-function-convolved, bulge--disk decompositions are performed in both the {\it r} and {\it g} bands using three different galaxy fitting models: a pure S\'ersic model, a bulge + disk model, and a S\'ersic with free index bulge + disk model (see Simard et al. 2011 for details). We chose to construct the most conservative sample of bulgeless galaxies identified by this catalog by requiring a bulge-to-total (B/T) ratio of 0 in both the {\it r} and {\it g} bands for the best-fit model.  We also required that SDSS spectroscopic data were available for all galaxies so that the redshifts of the sample were determined.  We cross-matched this SDSS bulgeless galaxy sample with the final {\it WISE} all-sky data release catlog\footnote{\url{http://wise2.ipac.caltech.edu/docs/release/allsky/}} to within $<$ $1\arcsec$, requiring that all galaxies be detected in the {\it WISE} $W1$ (3.4 $\mu$m), $W2$ (4.6 $\mu$m) , and  $W3$ (12 $\mu$m) bands with a signal-to-noise greater than $5\sigma$. The {\it WISE} catalog lists Vega magnitudes which are adopted throughout this paper. The cross-matched bulgeless galaxy sample obtained in this way contained 13,862 low-redshift galaxies. The redshift of the sample ranges from 0.002 to 0.33 with a mean of 0.08. The vast majority are nearby galaxies: $95\%$ have redshifts $< 0.15$. We obtained stellar masses and emission line fluxes for the galaxies in our sample from the Max Planck Institut fur Astrophysik/Johns Hopkins University (MPA/JHU) collaboration\footnote{\url{http://www.mpa-garching.mpg.de/SDSS/}}.  Of the 13,862 galaxies in our sample, $89\%$ had  H$\beta$, [OIII] $\lambda5007$, H$\alpha$, and [NII] $\lambda6584$ emission line fluxes detected with a signal to noise $> 3\sigma$.  For these galaxies, we determined their optical spectral classification using the [NII]/H$\alpha$ versus [OIII]/H$\beta$, line ratios using the widely-adopted BPT diagram (Baldwin, Phillips \& Terlevich 1981).  Galaxies were identified as AGNs based on their optical line ratios following the classification scheme of \citet{kewley2001}.  Based on this classification scheme, 57 galaxies, or $0.46\%$ of the sample, were classified as an AGNs. We also classified the galaxies in the sample using the demarcation between AGNs and star-forming galaxies from \citet{stasinska2006} since some of the galaxies classified as star-forming by \citet{kewley2001} may contain up to a 20\% contribution from an AGN (see \citet{stasinska2006}).   Using this classification scheme, 10,318~(83\%) of the galaxies were classified as star-forming galaxies.  Galaxies between these two extremes are often referred to as ``composite'' objects, which can include highly star-forming, galaxies harboring shock-excited gas, as well as galaxies that contain contributions to the emission lines  from both AGNs and HII regions. Only 14\% of the sample falls in this category. The stellar masses of the galaxies in the sample range from log(M/M$_{\odot}$) = 6.18 to 11.86, with a mean value of 10.02.

There are several important considerations we point out regarding the sample presented in this paper.  First, we emphasize that the ``bulgeless '' classification we employ is based on the automated bulge-disk decompositions from the \citet{simard2011} catalog, and represents galaxies that lack a well-fitted bulge. For simplicity, we will refer hereafter to these galaxies as bulgeless. The low resolution and sensitivity of the SDSS images precludes us from determining if there are any faint bulges.  To evaluate the robustness of the bulgeless classification in our assembled sample, we performed two dimensional bulge disk decompositions using version 3.2 of GIM2D and the identical procedure employed in  \citet{simard2011} on thousands of artificial simulated galaxies with true bulge fractions of B/T $>$ 0.025 and  {\it r} band half light radii and magnitudes similar to the galaxies in our sample in order to determine how many would be falsely classified as bulgeless.  Our bulgeless galaxy sample has a mean half light radius in the {\it r}-band of 5 pixels and 95\% of our galaxies are brighter than an {\it r}-band apparent magnitude of 17.5.  For these values, the probability of a galaxy with a B/T even as low as 0.025 being misclassified as having B/T $< $ 0.010 is 4\%, and drops to zero with B/T=0.1.  Therefore, the majority of the galaxies in the sample are not likely to have significant bulges based on the SDSS data.  Higher spatial resolution and more sensitive imaging observations, for example from the {\it Hubble Space Telescope}, are crucial to assess and characterize any faint bulge components and to obtain robust structural parameters for the host galaxies  in this sample.  

Second, it is important to consider whether any of the galaxies in the sample are interacting in order to determine if the galaxies are ``pristine'', with SMBHs that have evolved secularly over cosmic time.   We investigated the merger status of the galaxies in our sample using the Galaxy Zoo project\footnote{\url{http://www.galaxyzoo.org}}.  Galaxy Zoo users were asked to identify morphological signs of interactions by selecting a ``merger'' button.  Of the 13,862 galaxies in our bulgeless galaxy sample, 13,060 had Galaxy Zoo classifications available.  We used the weighted-merger-vote-fraction, $f_{m}$, to explore the interaction status of the sample.  This parameter varies from 0 to 1, where 0 represents a clearly isolated galaxy and a value of 1 represents a definitive merger \citep[for details, see][]{darg2010}. Based on an investigation of the Galaxy Zoo catalog, \citeauthor{darg2010} find that $f_{m} > 0.6$ robustly identifies mergers. For $f_{m} < 0.4$, misclassifications occur based on projection of galaxies and stars along the line of sight as well as cases where the galaxy hosts may appear disturbed but there are no visual signs of a companion. In the entire bulgeless sample with Galaxy Zoo classifications, 12,855 (98\%) have $f_{m} < 0.4$, 12,135 (93\%) have  $f_{m} < 0.1$ , and 9,176 (70\%) have  $f_{m}=0$, indicating a high probability that they are isolated galaxies.  Although the aforementioned cautions apply to the selection of bulgeless AGNs with close companions (i.e., in terms of their genesis and potential selection issues), we have not explicitly excluded these from our sample, as they may still be interesting for future study.  However, we emphasize that the vast majority of bulgeless galaxies appear to be isolated bulgeless galaxies based on SDSS data.

\section{Mid-IR Colors of Bulgeless Galaxies and {\it WISE} AGN Selection Criteria}

Extensive efforts over the past decade have demonstrated the power and reliability of mid-infrared observations in discovering optically hidden AGNs. This is thought to be because hot dust surrounding AGNs produces a strong mid-infrared continuum and infrared spectral energy distribution (SED) that is clearly distinguishable from normal star-forming galaxies for both obscured and unobscured AGNs in galaxies where the emission from the AGN dominates over the host galaxy emission \citep[e.g.][]{stern2012}.  In particular, at low redshift, the $W1-W2$ color of galaxies dominated by AGNs is considerably redder than that of inactive galaxies\citep[see Figures 1 in][]{stern2012,assef2012}.  At higher redshifts ( z $>$ 1.5) the host galaxy becomes red across these bands but becomes undetectable by {\it WISE}.  

 There are several {\it WISE} color diagnostics that have been employed in the literature to select AGNs.  Based on the {\it Spitzer-WISE} ecliptic pole data, which contains the highest coverage depths achieved by {\it WISE}, \citet{jarrett2011} using the first 3 {\it WISE} bands define an ``AGN'' region in  $W1-W2$ versus $W2-W3$ color-color space that separates  AGNs that dominate over their host galaxies from normal galaxies.   \citet{stern2012} show using the shallower COSMOS data that while inclusion of the longer wavelength bands increases the reliability of AGN selection due to contamination from high redshift (z$>$1.3) galaxies, for low redshifts, {\it WISE} is able to robustly identify AGNs using the more sensitive $W1$ and $W2$ bands alone.    \citet{donley2012} demonstrate using a combination of galaxy templates and real data from pure starbursts identified from {\it Spitzer} IRS spectroscopy that this mid-infrared color selection has minimal contamination from purely star-forming galaxies below a redshift of $z \sim 1$.  Using the IRAC-selected AGN as their ``truth sample'' \citet{stern2012} show that color cuts of $W1-W2>0.8$, $W1-W2>0.7$, $W1-W2>0.6$, $W1-W2>0.5$ identify IRAC AGNs with a reliability of 95\%, 85\%,70\%, and 50\%, respectively.  However, they point out that these numbers are conservative, since it is unknown whether or not the ``non-AGN'' contaminants are in fact really active. In fact, they show that some of the contaminants with spectroscopic redshifts available are indeed confirmed to be broad-line quasars. Given that the redshifts of the galaxies in our sample are low, and that {\it WISE} color selections are critically sensitive to the AGN's power  relative to that of the host galaxy,  we explore a range of these published color cuts in our bulgeless sample.
  
\begin{figure}
\noindent{\includegraphics[width=8.7cm]{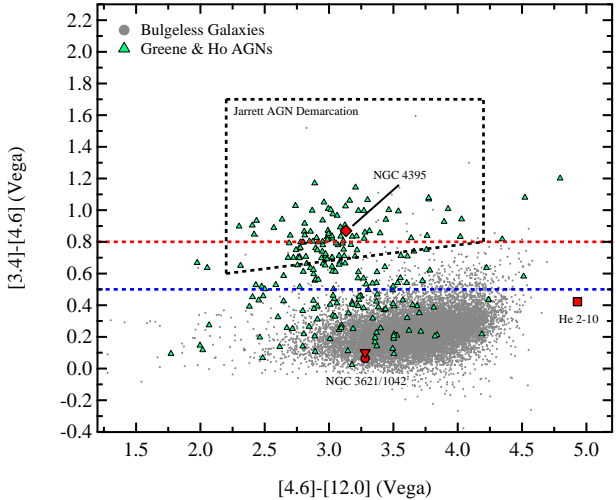}}
\caption{$W1-W2$ color versus the  $W2-W3$ color for our bulgeless galaxy sample. The AGN region from \citet{jarrett2011} is shown together with the$W1-W2> 0.8$ (red dashed line) and $W1-W2>0.5$ (blue dashed line) color cuts. Several known bulgeless galaxies with confirmed AGNs are highlighted along with those from the sample of low mass optically identified broad line AGNs  from \citet{greene2007}\\}
\label{colorcolorplot}
\end{figure}

In Figure~\ref{colorcolorplot}, we plot the $W1-W2$ versus $W2-W3$ color for the galaxies in our bulgeless sample together with those from a few purely bulgeless galaxies with AGNs from the literature and those from the sample of broad line AGNs with low mass black holes from \citet{greene2007}.  The AGN region from \citet{jarrett2011} is shown together with the  $W1-W2 >0.8$ and $W1-W2 >0.5$ color cuts.  In our bulgeless galaxy sample, there are 30 galaxies that fall within the \citet{jarrett2011} AGN region, 25 with  $W1-W2 >0.8$, 51 with  $W1-W2 >0.7$ , 101 with  $W1-W2 >0.6$ , and 353 with  $W1-W2 >0.5$ .  We  note that even the least stringent color cut ( $W1-W2 >0.5$) identifies galaxies in which the AGN emission dominates over the host galaxy emission.  In galaxies with weak AGNs, dilution of the mid-infrared AGN continuum by the host galaxy light can cause bluer $W1-W2$ colors that become indistinguishable from star forming galaxies (see Figure 1 in \citet{stern2012}). In fact, a large fraction of the optically identified AGNs in the entire SDSS sample in this redshift range have $W1-W2$ color below this cutoff  \citep[e.g.][]{Yan2012}. Any of the {\it WISE} color cuts shown in  Figure~\ref{colorcolorplot} therefore only identifies AGNs that dominate the emission from their host galaxies. Indeed, as can be seen from Figure~\ref{colorcolorplot},  the bulgeless galaxies NGC 3621, NGC 1042, and He 2-10, all discovered recently to host AGNs, all have $W1-W2$ colors below those of any of the color cuts adopted in this work since the AGNs are weak compared to the host galaxy.  Even within the sample from  \citeauthor{greene2007}, which are all indisputable broad line AGNs, 72\% have  $W1-W2 < 0.8$, and 36\% have $W1-W2 < 0.5$ .  

Since the redshifts of our bulgeless galaxy sample are low, well below the threshold at which high redshift contaminants are a concern, and all of the color cuts shown in Figure~\ref{colorcolorplot} identify AGNs that dominate over their host galaxy emission, it is possible that a significant fraction of the bulgeless galaxies with $W1-W2 >0.5$ do indeed host AGNs, and therefore these galaxies should be investigated with follow-up multi wavelength observations.  

\section{Properties of Bulgeless {\it WISE}-Identified AGN Candidates}

In Table 1, we list the {\it WISE} $W1-W2$ colors, masses, and redshifts of the bulgeless galaxies that meet the most stringent three-band color cut from \citet{jarrett2011}.  These are the AGN candidates most likely to contain dominant AGNs based on their mid-infrared colors. All galaxies are nearby, with redshifts ranging from 0.016 to 0.27, with a mean redshift of 0.13.  The masses of the galaxies range from log(M/M$_{\sun}$) $=$ 8.1 to 11.3, with a mean of 10.5.

\begin{table}
\caption{ Most Extreme AGN Candidates Identified using {\it WISE} 3-band color cut }

\begin{tabular}{lccc}
\hline
\hline
\noalign{\smallskip}
SDSS Name & W1-W2 (mag.) & log($M/M_{\sun}$) & $z$  \\

\noalign{\smallskip}       
\hline
\noalign{\smallskip}

J012218.11+010025.8	&	1.60	&	10.0	&	0.055	\\
\noalign{\smallskip}
J131851+340124.9	&	1.52	&	10.0	&	0.056	\\
\noalign{\smallskip}
J103631.87+022144	&	1.30	&	10.5	&	0.050	\\
\noalign{\smallskip}
J155409.08+145703.5	&	1.14	&	10.9	&	0.137	\\
\noalign{\smallskip}
J142231.37+260205	&	1.11	&	10.7	&	0.159	\\
\noalign{\smallskip}
J122434.66+555522.3	&	1.10	&	10.3	&	0.052	\\
\noalign{\smallskip}
J075039.96+292845.7	&	1.00	&	10.4	&	0.147	\\
\noalign{\smallskip}
J123304.57+002347.1	&	0.98	&	10.0	&	0.069	\\
\noalign{\smallskip}
J143644.55+024715.8	&	0.97	&	11.3	&	0.221	\\
\noalign{\smallskip}
J112824.27+115615.2	&	0.94	&	10.4	&	0.162	\\
\noalign{\smallskip}
J142847.54+324436.8	&	0.93	&	10.8	&	0.195	\\
\noalign{\smallskip}
J232020.09+150420.5	&	0.91	&	10.6	&	0.148	\\
\noalign{\smallskip}
J130131.53+212748.7	&	0.91	&	10.1	&	0.087	\\
\noalign{\smallskip}
J033331.86+010717.1	&	0.87	&	10.7	&	0.178	\\
\noalign{\smallskip}
J092907.78+002637.2~$^a$	&	0.85	&	11.0	&	0.117	\\
\noalign{\smallskip}
J103549.57+122406.7	&	0.84	&	10.4	&	0.168	\\
\noalign{\smallskip}
J132932.41+323417	&	0.84	&	8.1	&	0.016	\\
\noalign{\smallskip}
J110413.07+301014.1	&	0.82	&	10.2	&	0.120	\\
\noalign{\smallskip}
J135302.41+082220.9	&	0.81	&	11.3	&	0.267	\\
\noalign{\smallskip}
J085121.74+094010.2	&	0.81	&	10.5	&	0.112	\\
\noalign{\smallskip}
J114753.62+094551.9~$^a$	&	0.81	&	10.3	&	0.095	\\
\noalign{\smallskip}
J140515.45+581247.3~$^a$	&	0.77	&	11.3	&	0.250	\\
\noalign{\smallskip}
J085236.35+292853.2~$^a$	&	0.76	&	10.2	&	0.079	\\
\noalign{\smallskip}
J103222.94+361727.8	&	0.74	&	10.9	&	0.236	\\
\noalign{\smallskip}
J155448.29+090817.9	&	0.74	&	10.9	&	0.166	\\
\noalign{\smallskip}
J133341.72+653619.1~$^a$	&	0.73	&	10.9	&	0.158	\\
\noalign{\smallskip}
J132647.54+101436.4	&	0.73	&	10.3	&	0.089	\\
\noalign{\smallskip}
J122809.2+581431.5	&	0.72	&	10.5	&	0.110	\\
\noalign{\smallskip}
J085153.64+392611.8	&	0.71	&	10.6	&	0.130	\\
\noalign{\smallskip}
J124820.31+181354.2~$^a$	&	0.71	&	10.8	&	0.110	\\
\noalign{\smallskip}
\hline

\noalign{\smallskip}
${\rm ^a}$ Optically-identified AGNs & & & \\

\end{tabular}

\end{table}

\begin{figure*}[]

\centering

\noindent{\includegraphics[width=16cm]{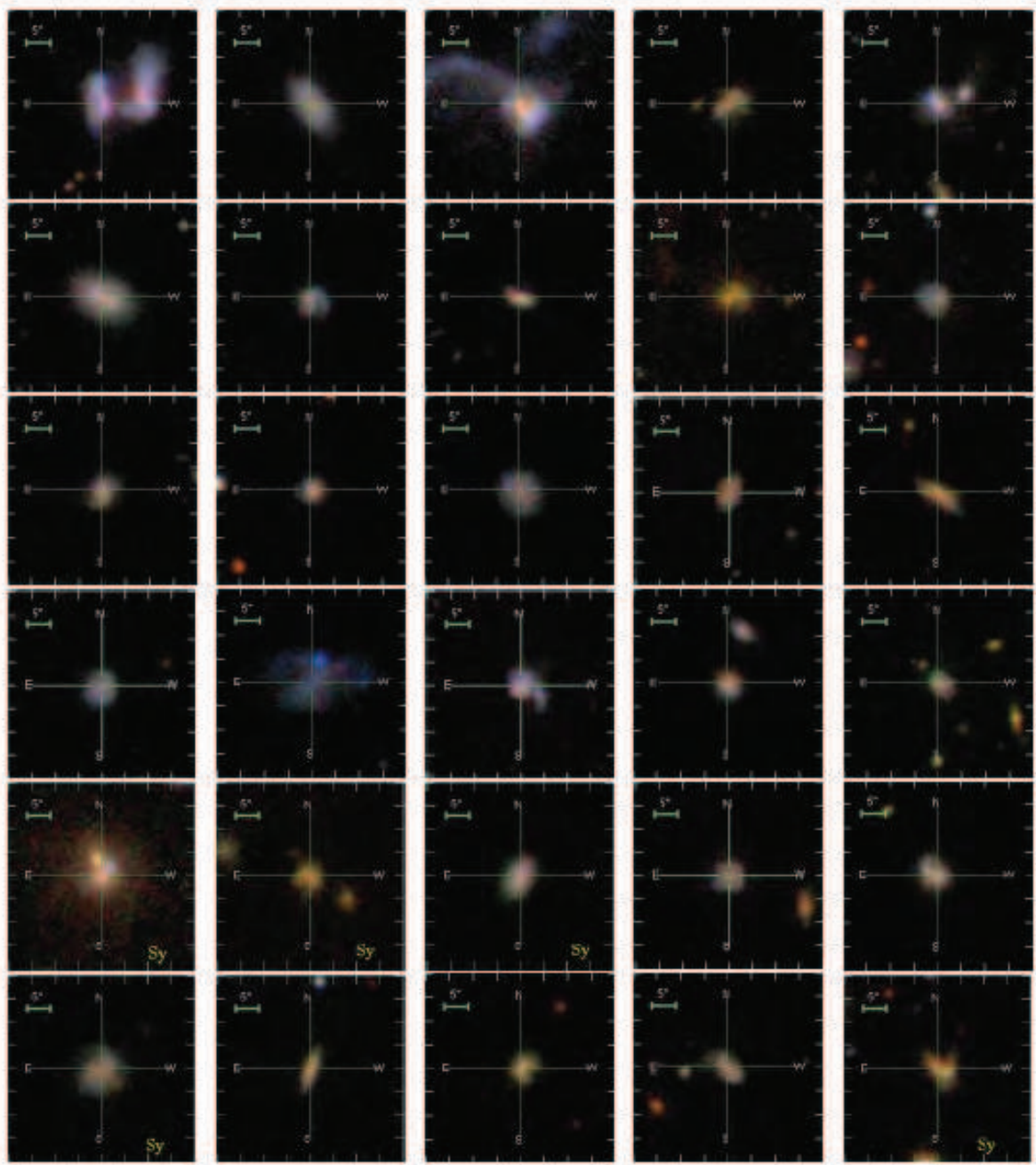}}

\caption{\footnotesize{SDSS images of AGN candidates identified by the 3-band {\it WISE} color diagnostics from \citet{jarrett2011} arranged in order of decreasing $W1-W2$ color .  Galaxies with classified as Seyferts by their optical spectra are indicated as ''Sy'' .}}
\label{sdssimages}
\end{figure*}

\begin{figure}
\noindent{\includegraphics[width=8.7cm]{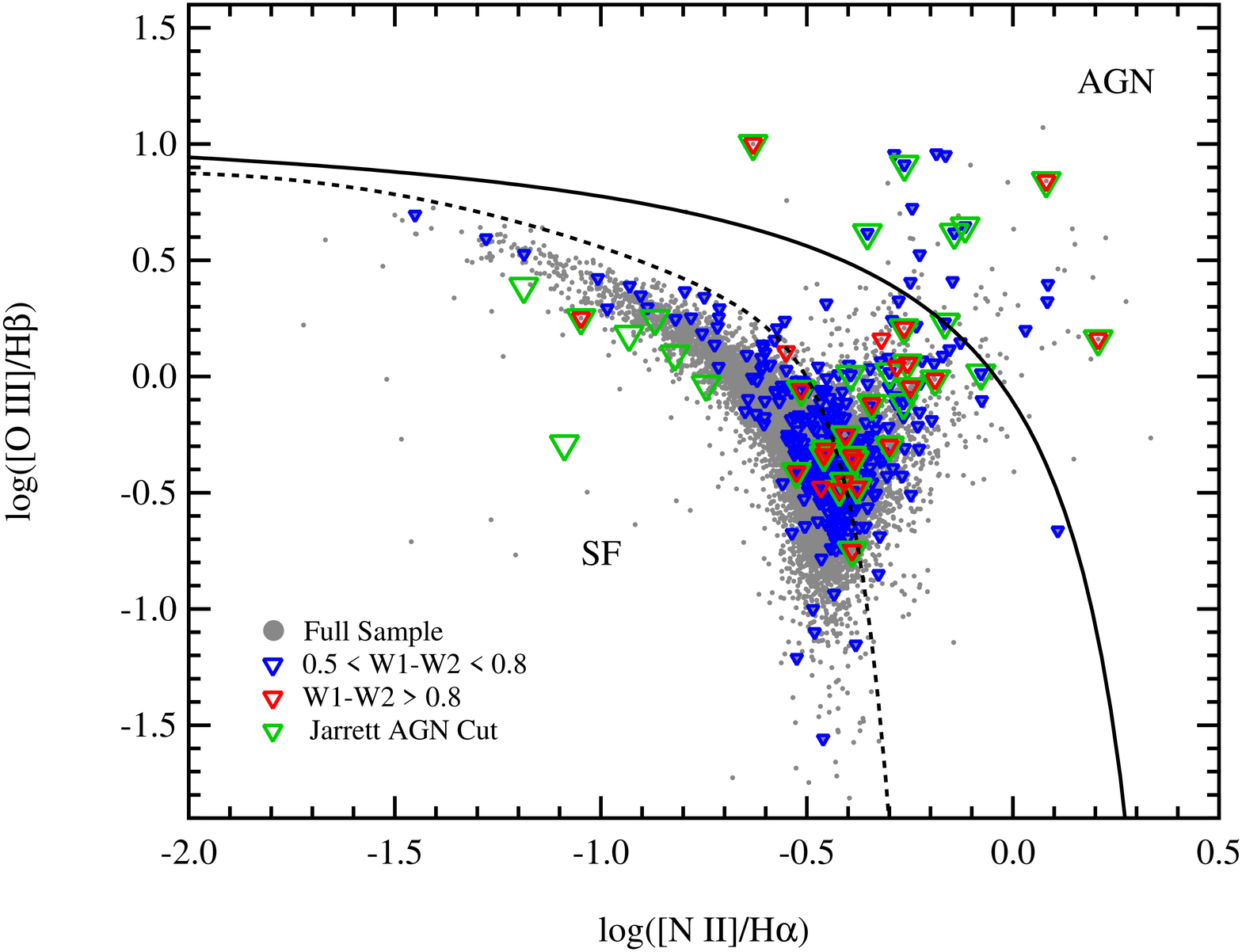}}
\caption{BPT diagram of the bulgeless galaxies from SDSS.  Galaxies with mid-IR color consistent with various mid-IR color cuts (from \citet{jarrett2012}, $W1-W2 > 0.8$,$W1-W2 > 0.5$ ) are highlighted. The black lines correspond to the demarcations separating star forming galaxies from AGNs from  \citet{stasinska2006} and \citet{kewley2001}. \\}
\label{opticalbpt}
\end{figure}

 In Figure~\ref{sdssimages}, we display the SDSS images of these AGN candidates organized by decreasing $W1-W2$ color, as listed in Table 1.   The morphologies of the {\it WISE} AGN hosts range from low surface brightness objects with little structure to more structured bulgeless disks, and some are clearly interacting galaxies. Galaxies with optical line ratios consistent with AGNs are indicated in the images.  The majority of {\it WISE}-identified AGN candidates in Table 1 show no signatures of an AGN based on their optical spectra, suggesting that this population is obscured in the optical.  In Figure~\ref{opticalbpt}, we plot the optical BPT diagram for all galaxies in our bulgeless galaxy sample with the optical emission line fluxes detected above the $3\sigma$ level.  Galaxies with {\it WISE} colors consistent with AGNs based on the various  {\it WISE} color cuts are highlighted. Of the 25 galaxies that meet the $W1-W2>0.8$ color cut, only 2 have optical line ratios that are above the \citet{kewley2001} AGN region of the BPT region. Only 4\% of the galaxies that have $W1-W2>0.5$ are in the  \citet{kewley2001} AGN region of the BPT diagram. In a complementary analysis exploring black hole scaling relations, \citet{marleau2013} have also identified a population of {\it WISE} AGNs, and explored their distribution of bulge and disk masses.

\section{Alternative Scenarios and Caveats}

While the red mid-infrared color of the bulgeless galaxies presented in this work is highly suggestive of AGN activity, it is possible that the red colors are due solely to star formation. For example, ultraluminous infrared galaxies (ULIRGs) can have $W1-W2$ colors well in excess of 1~\citep[e.g.][]{jarrett2011,Yan2012}, although these red colors may indeed be due to AGN activity since many if not most ULIRGs host obscured AGNs~\citep[e.g.][]{veilleux2009}. We searched for {\it IRAS} detections for the bulgeless AGN candidates.   Only 5 of the 30 galaxies listed in Table 1  were listed in the  in the {\it IRAS} Faint Source Catalog, and one (SDSS J142231.37+260205.0) is a ULIRG.  Of the 353 bulgeless galaxies that meet the $W1-W2 > 0.5$ color cut, 54 were listed in the {\it IRAS} Faint Source Catalog, and only 7 of these are ULIRGs.  Most (87\%) of the galaxies were not detected by {\it IRAS} in 3 or more bands. The red {\it WISE} colors of the majority of the galaxies in our sample therefore cannot be attributed to extreme infrared luminosities, due either to an intense obscured starburst or an extremely luminous AGN in a ULIRG host.

\begin{figure}
\noindent{\includegraphics[width=8.7cm]{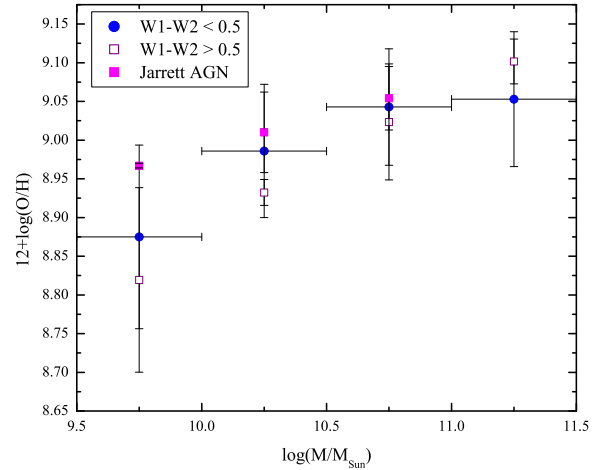}}
\caption{Mean Oxygen abundance 12+log(O/H) of the {\it WISE} AGN candidates using the various color cuts explored in this work compared with galaxies in the bulgeless sample with $W1-W2<0.5$.  \\}
\label{metallicity}
\end{figure}

In addition to ULIRGs, there have been a handful of low metallicity blue compact dwarfs (BCDs) with extreme ($W1-W2>2$) mid-infrared colors, raising the possibility that there is a similar origin for the hot dust in the bulgeless galaxy sample. Since the hardness of the stellar radiation increases with decreasing metallicity \citep[e.g.][]{campbell1986}, and BCDs contain significant star formation, the dust in BCDs can potentially be heated to higher temperatures than is typically seen in starburst galaxies.   Given that many of the galaxies in our sample are low mass galaxies, we explored the possibility that the red colors in our sample have an origin similar to the handful of BCDs described above.  Of the 13,862 galaxies in the bulgeless sample, 10,324 (74\%) had metallicities available from the MPA/JHU catalog.\footnote{Since the majority of {\it WISE} AGNs are optically classified as star--forming, strong line metallicity diagnostics are assumed to be applicable.}  Of the 353 bulgeless galaxies with $W1-W2 > 0.5$, 248 had metallicities available. In Figure~\ref{metallicity}, we plot the average metallicity of the galaxies in several mass bins in our sample with red mid-infrared colors using the various color cuts compared with those with $W1-W2<0.5$. As can be seen, the metalicities  of the red {\it WISE} galaxies are at or above solar, and there is no significant difference between the metalicities of the red sources and those of the galaxies with $W1-W2<0.5$ in the sample. 

Furthermore, if the red {\it WISE} colors in our sample are purely a metallicity effect, one would expect a correlation between the mid-infrared color and metallicity, with redder colors corresponding to lower metallicity galaxies. However, we find no correlation between the $W1-W2$ color and metallicity in our sample,  strongly suggesting that metallicity effects are not responsible for the red colors, at least for the vast majority of galaxies in our sample.

It is also worth mentioning that while star formation is often cited as the origin of the hot dust emission in BCDs, it is not clear whether in all cases it is simply a by-product of the relatively hard radiation field produced by extreme star formation in low metallicity environments, or if AGNs could play a role.  Indeed, in some BCDs, high ionization emission lines generally associated with AGNs have been seen \citep[e.g.][]{Izotov2004,Izotov2012, fricke2001,Izotov2001}, including in the galaxy SBS 0335Ð052E. The detection of these high ionization lines implies the presence of a hard radiation field which cannot be produced by models of high-mass X-ray binaries or massive stellar populations even at low metallicities, but instead requires an AGN and/or fast radiative shocks \citep{Izotov2012}. \citet{Izotov2008} also found four low-metallicity dwarf galaxies with very broad H$\alpha$ emission requiring the presence of an accretion disk around an intermediate-mass black hole. The presence of an AGN in at least some of the BCDs with red {\it WISE} colors has not been ruled out.

\section{Growing Evidence of SMBHs in Bulgeless and Low Mass Galaxies}

The preliminary analysis above suggests that the red {\it WISE} colors discovered in the bulgeless galaxies presented in this work cannot easily be explained by any non-accretion based mechanism for the majority of galaxies.  Confirmation of AGN activity requires follow-up high spatial resolution X-ray or infrared spectroscopic observations.  We searched both the {\it XMM Newton} and {\it Chandra} archives for observations of the $W1-W2 \geq 0.5$ galaxies in our sample.  Of the 353 galaxies, we found observations  for only 3 galaxies in the {\it Chandra} archive.  Of these observations, two were relatively short at 8.5 ks and 2.0 ks, and unsurprisingly there were no nuclear  X-ray detections.  The $3\sigma$ upper limits for the X-ray luminosity of these sources are large ( $L_{0.3-8\rm{~keV}}<4\times10^{42}$ ergs s$^{-1}$ and $L_{0.3-8\rm{~keV}}<2\times10^{42}$ ergs s$^{-1}$, respectively) and therefore do not rule out the possibility that these galaxies contain AGNs.  The third source, IRAS F12112+0305, observed for 10 ks, shows a compact, double X-ray nucleus associated with a galaxy that is clearly in the stages of a late merger.  The physical separation of these X-ray peaks is 3.7 kpc, and the combined hard X-ray luminosity is $L_{2-10\rm{~keV}}=3.8\times10^{41}$ ergs s$^{-1}$~\citep{teng2005}.  The infrared colors of this source is $W1-W2=0.82$ and, perhaps not coincidentally, this source is classified as a ULIRG.  It falls just outside of the \citet{jarrett2011}``AGN'' region in infrared color space due to its red $W2-W3$ color.  While this source is not typical of our sample, it is the one galaxy in our sample with X-ray observations, and the X-ray observations confirm the AGN. Unfortunately, none of the low-mass sources in our sample have archival X-ray observations, since they are not generally the targets of X-ray proposals.  

We recently acquired {\it XMM-Newton} observations of 3 nearby bulgeless galaxies with red colors.  Two of the 3 galaxies show nuclear X-ray point sources consistent with AGNs.  Remarkably, the most X-ray luminous of these two galaxies is a bulgeless irregular optically quiescent dwarf galaxy similar in morphology to He 2-10, but a factor of ten less massive and a factor of ten more luminous in the X-rays, making it the lowest mass dwarf with an AGN currently known \citep{secrest2013}.  These preliminary results strongly suggest that {\it WISE} is uncovering a population of obscured AGNs that dominate over their host galaxies in the low mass regime.
Finally, we also conducted a preliminary  search of the {\it VLA} archive for extreme {\it WISE} galaxies selected such that $W1-W2 \geq 0.8$.  We found one pair of merging galaxies with archival 1.4 GHz radio data (J012217+010027). The radio data show two compact sources coincident with the apparent $r$-band peaks of the two galaxies, suggesting the presence of AGNs, combined with extended radio emission coincident with material bridging the gap between the galaxies~\citep{hodge2011}.  

This work adds to the growing evidence that SMBHs are found in bulgeless and low mass galaxies. There is incontrovertible X-ray and radio evidence for an AGN in the dwarf galaxy He 2-10~\citep{reines2011}, and over a hundred dwarf galaxies have recently been discovered with optical spectroscopic signatures of AGNs ~\citep{2013ApJ...775..116R} .   \citet{simmons2013} have identified a population of optically identified AGNs in bulgeless galaxies.  \citet{schramm2013} recently reported the discovery of 3 AGNs in low mass galaxies in the {\it Chandra} Deep Field South.  One of the 3 galaxies was detected by {\it WISE} and has a $W1-W2$ color of 0.48, just below the lowest color cut discussed in this work.  The discovery of AGN activity in bulgeless and low mass galaxies is therefore  not unprecedented. Furthermore, there are 57 optically-identified AGNs in our bulgeless sample, 16 of which also have $W1-W2$ greater than 0.5, suggesting that the AGN dominates over the host galaxy emission in these galaxies.  In AGN-dominated galaxies, the $W2$ luminosity is expected to originate almost entirely from the AGN~\citep{stern2012}.  In the 16 optical AGNs that also have red {\it WISE} colors, the $W2$ luminosities range from log($L_{W2}$ (ergs s$^{-1})$) $=42.3$ to 44.6, with a mean of 43.7.  By comparison,the $W2$ luminosities of  our full sample of 337 optically-normal galaxies with $W1-W2\geq0.5$ range from log($L_{W2}$ (ergs s$^{-1})$)$=41.6$ to 45.0, with a mean of 43.8 (ergs s$^{-1}$).  Thus, even if a putative AGN in these galaxies dominates over its host, the implied total AGN luminosity is by no means extreme.  

Based on all available data and the published work on {\it WISE} colors of galaxies thus far, the red mid-infrared colors of the bulgeless galaxies presented in this work is difficult to explain without invoking the presence of obscured AGNs. Follow-up multi wavelength investigations of the full $W1-W2 >0.5$ should be conducted.  If these galaxies are found out not to host obscured AGNs, we emphasize that they are extremely unusual, and that the origin of the host dust needs to be understood in this population.

\section{Comparison of {\it WISE} and optical AGN selection}

\begin{figure}
\noindent{\includegraphics[width=8.7cm]{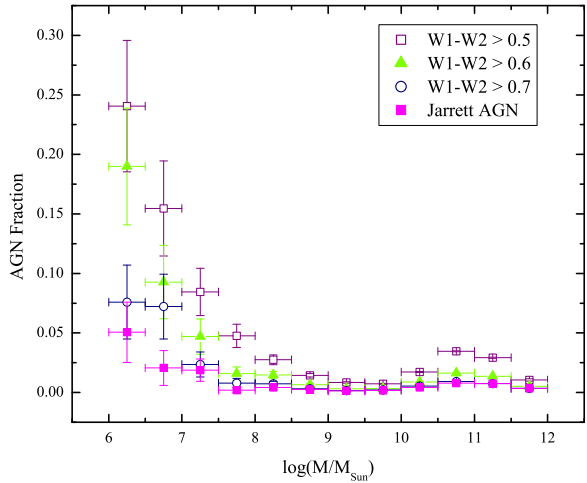}}
\caption{Fraction of AGN candidates identified by  {\it WISE} as a function of stellar mass in the cross--matched {\it WISE}/ \citet{simard2011} catalogs. \\}
\label{agnfractionvsmass}
\end{figure}

If the red mid-infrared colors of the galaxies in our sample are indeed due to AGN activity, our investigation has revealed a  completely unexpected population in optically quiescent bulgeless galaxies. It is well known that the AGN fraction based on optical studies is highest in massive bulge-dominated galaxies \citep{kauffmann2003}, and drops dramatically in the low mass regime. While the fraction of bulgeless galaxies that host mid-infrared selected AGNs is still low, if a significant fraction of the galaxies with $W1-W2>0.5$ are confirmed to host AGNs, the AGN detection rate based on {\it WISE} could be a factor of up to  $\approx$ 6 larger than the optical AGN detection rate, at least in galaxies that are not massive and bulge-dominated. In order to compare the general dependence of {\it WISE} AGN selection with  optical AGN selection as a function of stellar mass, we obtained the {\it WISE} colors for all galaxies in the \citet{simard2011} catalog for all host morphologies. Again we crossmatched the \cite{simard2011} catalog with the {\it WISE} all-sky data release catalog to within $<1\arcsec$, requiring a detection threshold above $5\sigma$ in the  $W1$, $W2$, and $W3$ bands.  In Figure~\ref{agnfractionvsmass}, we show the fraction of galaxies classified as AGN by {\it WISE} according to the various color-cuts as a function of stellar mass.   Although the reliability of AGN selection is a function of the adopted color cut, Figure~\ref{agnfractionvsmass} clearly shows that the AGN fraction based on infrared color selection is highest at the lowest stellar masses and drops dramatically at higher stellar masses.  This result is in direct contrast with results from optical spectroscopic surveys.  In the \citet{kauffmann2003} study of 22,623 narrow-line AGNs drawn from SDSS the AGN fraction approaches nearly 100\% for all emission line galaxies at the highest masses and drops dramatically with decreasing stellar mass (see Figure 5, solid histogram in top panel, in \citet{kauffmann2003}).  A key and striking result from this study is that {\it WISE} reveals a population of optically-hidden AGNs in the lowest mass galaxies, and that the incidence of {\it WISE} AGNs remarkably follows an opposite trend with mass than is found in optical spectroscopic studies. We note too that as discussed above, {\it WISE} detects AGNs in galaxies where the AGN dominates over the host galaxy emission.  Indeed, as can be seen from Figure~\ref{agnfractionvsmass}, none of the {\it WISE} color cuts  select the bulk of the optically identified AGNs at higher masses, since the AGNs don't dominate over the host galaxy emission.  Mid-infrared color selection is therefore selecting a more extreme population of AGNs that is more prevalent in the low mass regime.

\section{Summary}

We have conducted an investigation of the {\it WISE} colors of 13,862 low redshift galaxies drawn from and classified as bulgeless from the SDSS catalog of bulge/disk decompositions from \citet{simard2011}.  Our main results can be summarized as follows:
 
 \begin{enumerate}
\item{We have identified a population   bulgeless galaxies that display extremely red  mid-infrared color indicative of hot dust  highly suggestive of a dominant active galactic nucleus (AGN). Depending on the {\it WISE} color selection employed, the number of AGN candidates ranges from 25 to over 300. }
\item{The vast majority of these galaxies display no optical AGN signatures in their SDSS spectra, and many appear to be ``pristine'' isolated galaxies with no signs of interactions.}
\item{Based on existing data, we have explored other known scenarios that can heat the dust to high temperatures including extreme star formation and star formation in low metallicity environments, but none of these scenarios appear to explain the observed colors of this sample. }
\item{ If these galaxies do host AGNs, the AGN detection rate can be up to a factor of $\approx$ 6 larger than the optical AGN detection rate in this host galaxy demographic. }
\item{ We compare the AGN detection rate based on infrared color selection with the optical AGN detection and find that the fraction of all galaxies identified as candidate AGNs by WISE is highest at the lowest stellar masses and drops dramatically in the highest mass galaxies, in striking contrast to the findings from optical studies.}
\end{enumerate}

The discovery of optically quiescent bulgeless  with AGNs  will add to the growing evidence that SMBH growth takes place through secular processes and can potentially help discriminate between scenarios for the seed population from which they arose.  Follow-up multiwavelength high spatial resolution observations of this sample are critical to confirm the presence of AGNs in these galaxies, to obtain more precise estimates of the bulge masses, and to constrain the black hole masses, accretion rates, and nuclear bolometric luminosities. If , on the other hand, these galaxies do not turn out to host AGNs, they are very unusual and the origin of their hot dust is of interest in itself.  This scenario would also have an impact on the use and reliability of {\it WISE} color selection of AGN in other surveys.

\acknowledgements
We thank the referee for a very thorough and insightful review, which considerably improved the quality of the paper. We are very grateful to Roberto Assef for helpful discussions as well as running his stellar templates over the redshift range of this sample to confirm that the adopted {\it WISE} color cut used in this paper is significantly higher than that of any star forming galaxy template in our redshift range. This paper benefited greatly from discussions and advice on bulgeless galaxy selection with Luc Simard, who also provided the results from his own simulations to enable us to evaluate the robustness of the bulgeless classification of our sample. It is a pleasure to thank Mario Gliozzi for numerous discussions and collaborations on parallel projects that helped motivate this work. We are also very grateful to Luis Ho and Aaron Barth for their insightful comments on this work.  Basic research in IR astronomy at NRL is funded by the US ONR. This publication makes use of data products from the {\it Wide-field Infrared Survey Explorer}, which is a joint project of the University of California, Los Angeles, and the Jet Propulsion Laboratory/California Institute of Technology, funded by the National Aeronautics and Space Administration. We are grateful to the MPA/JHU group for access to their data products and catalogues (maintained by Jarle Brinchmann at \url{http://www.mpa-garching.mpg.de/SDSS/}). Funding for the SDSS and SDSS--II has been provided by the Alfred P. Sloan Foundation, the Participating Institutions, the National Science Foundation, the U.S. Department of Energy, the National Aeronautics and Space Administration, the Japanese Monbukagakusho, the Max Planck Society, and the Higher Education Funding Council for England. The SDSS Web Site is \url{http://www.sdss.org/}. The SDSS is managed by the Astrophysical Research Consortium for the Participating Institutions. The Participating Institutions are the American Museum of Natural History, Astrophysical Institute Potsdam, University of Basel, University of Cambridge, Case Western Reserve University, University of Chicago, Drexel University, Fermilab, the Institute for Advanced Study, the Japan Participation Group, Johns Hopkins University, the Joint Institute for Nuclear Astrophysics, the Kavli Institute for Particle Astrophysics and Cosmology, the Korean Scientist Group, and the Chinese Academy.


\begin{thebibliography}{}

\bibitem[Abel \& Satyapal (2008)]{abel2008}
Abel, N.~P., \& Satyapal, S.~2008, \apj, 678, 686

\bibitem[Araya Salvo et al.~(2012)]{araya2012}
Araya Salvo, C., Mathur, S., Ghosh, H., Fiore, F., \& Ferrarese, L.~2010, \apj, 757,179

\bibitem[Assef et al.~(2010)]{assef2010}
Assef, R.~J., et~al.~2010, \apj, 713, 970

\bibitem[Assef et al.~(2012)]{assef2012}
Assef, R. et al.~2012, \apj, in press

\bibitem[Baldwin, Phillips \& Terlevich (1981)]{baldwin1981}
Baldwin, J.~A., Phillips, M.~M., Terlevich, R.~1981, \pasp, 93, 5

\bibitem[Barth et al.~(2004)]{barth2004}
Barth, A.~J., Ho, L.~C., Rutledge, R.~E., \& Sargent, W.~L.~W.~2004, \apj, 607, 90

\bibitem[Bell et al.~(2003)]{Bell2003}
Bell, E.~F., McIntosh, D.~H., Katz, N., \& Weinberg, D.~2003, \apjs, 149, 289

\bibitem[Campbell et al.~(1986)]{campbell1986}
Campbell, A., Terlevich, R., \& Melnick, J.~1986, \mnras, 223, 811

\bibitem[Charlot \& Fall (2000)]{CharlotFall2000}
Charlot, S., \& Fall, M.~S.~2000, \apj, 539, 718C

\bibitem[Darg et al.~(2010)]{darg2010}
Darg, D.~W., et al.~2010, \mnras, 401, 1043

\bibitem[Desroches \& Ho (2009)]{desroches2009}
Desroches, L.~B., \& Ho, L.~C.~2009, \apj, 690, 267

\bibitem[Devriendt et al.~(1999)]{devriendt1999}
Devriendt, J.E.G., Guiderdoni, B., \& Sadat, R..~1999, A\&A, 350, 381

\bibitem[Dewangan et al.~(2005)]{dewangan2005}
Dewangan, G.~C., Griffiths, R.~E., Choudhury, M., Miyaji, T., \& Schurch, N.~J.~2005, \apj, 635, 198  

\bibitem[Donley al.~(2012)]{donley2012}
Donley, J.~L., et al.~2012, \apj, 748, 22  

\bibitem[Ellison et al.~(2011)]{ellison2011}
Ellsion, S.~L., Patton, D.~R., Mendel, J.~T., \& Scuder, J.~M.~2011, \mnras, 418, 2043

\bibitem[Ferrarese \& Merritt (2000)]{ferrarese2000}
Ferrarese, L., \& Merritt, D.~2000, \apj, 539, 9

\bibitem[Filippenko \& Ho (2003)]{filippenko2003}
Filippenko, A., \& Ho, L.~2000, \apj, 588, L13

\bibitem[Fricke et al.~(2001)]{fricke2001}
Fricke K.~J., Izotov Y.~I., Papaderos P., Guseva N.~G., \& Thuan T.~X.~2001, \aj, 121, 169 

\bibitem[Gebhardt et al.~(2001)]{gebhardt2001}
Gebhardt, K., et al.~2001, \apj, 555, 75

\bibitem[Gliozzi et al.~(2009)]{gliozzi2009}
Gliozzi, M., Satyapal, S., Erancleous, M., Titarchuk, L., \& Cheung, C.~C.~2009, \apj, 700, 1759

\bibitem[Ghosh et al.~(2008)]{ghosh2008}
Ghosh, H., Mathur, S., Fiore, F., \& Ferrarese, L.~2008, AIPC, 1053, 39G

\bibitem[Greene \& Ho (2007)]{greene2007}
Greene, J. \& Ho, L. 2007, \apj, 670, 92

\bibitem[Griffith et al.~(2011)]{griffith2011}
Griffith, R.~L., et al.~2011, \apj, 736, L22

\bibitem[Hodge et al.~(2011)]{hodge2011}
Hodge, J.~A., Becker, R.~H., White, R.~L., Richards, G.~T., \& Zeimann, G.~R.~2011, \apj, 142, 3

\bibitem[Houck et al.~(2004)]{Houck2004}
Houck, J.~R., Charmandaris, V., \& Brandl, B.~R.~2004, \apjs, 154, 211

\bibitem[Izotov et al.~(2001)]{Izotov2001}
Izotov Y.~I., Chaffee F.~H., \& Schaerer D.~2001, \aap, 378L, 45I

\bibitem[Izotov et al.~(2004)]{Izotov2004}
Izotov, Y.~I., et al.~2004, \aap, 415, L27

\bibitem[Isotive \& Thuan (2008)]{Izotov2008}
Izotov, Y.~I., \& Thuan, T.~X.~2008, \apj, 687, 133

\bibitem[Izotov et al.~(2009)]{Izotov2009}
Izotov, Y.~I., Guseva, N.~G., Fricke, K.~J., \& Papaderos, P.~2009, \aap, 503, 61

\bibitem[Izotov et al.~(2011)]{Izotov2011}
Izotov, Y.~I., Guseva, N.~G., Fricke, K.~J., \& Henkel, C. ~2011, \aap, 536, L7

\bibitem[Izotov, Thuan, \& Privan (2012)]{Izotov2012}
Izotov,Y.~I.,  Thuan, T.~X., \& Privon, G.~ 2012, \mnras, 427,1229

\bibitem[Jarrett et al.~(2011)]{jarrett2011}
Jarrett, T.~H. et al.~2011, \apj, 735,112

\bibitem[Jiang et al.~(2011)]{jiang2011}
Jiang, Y., Greene, J., Ho, L.~C., Xiao, T., \& Barth, A.~J.~2011, \apj, 742, 62

\bibitem[Jordi, Grebel, \& Ammon (2006)]{jordi2006}
Jordi, K., Grebel, E.~K., \& Ammon, K.~2006, \aap, 460, 339J

\bibitem[Kauffmann \& Haehnelt (2000)]{kauffmann2000}
Kauffmann, G., \& Haehnelt, M.~2000, \mnras, 311, 576

\bibitem[Kauffmann et al.~(2003)]{kauffmann2003}
Kauffmann, G., et al.~2003, \mnras, 346, 1055

\bibitem[Kennicutt (1998)]{Kennicutt1998}
Kennicutt, R.~C.~1998, \apj, 1998, 498, 541K

\bibitem[Kewley et al.~(2001)]{kewley2001}
Kewley, L.~J., Heisler, C.~A., Dopita, M.~A., \& Lumsden, S.~2001, \apjs, 132, 37

\bibitem[Lake et al.~(2012)]{Lake2012}
Lake, S.~E., et al.~2012, \apj, 143, 7

\bibitem[Marleau et al.~(2013)]{marleau2013}
Marleau, F., et al.~\mnras, submitted

\bibitem[McAlpine et al.~(2011)]{mcalpine2011}
McAlpine, W., Satyapal, S., Gliozzi, M., et al.~2010, \apj, 728, 1

\bibitem[Papaderos et al.~(1998)]{papaderos1998}
Papaderos, P., Izotov, Y.~I., Fricke, K.~J., Thuan, T.~X., \& Guseva, N.~G. 1998, \aap, 338, 43

\bibitem[Pustilnik, Pramskij, \& Kniazev (2004)]{pulstilnik2004}
Pustilnik, S.~A., Pramskij, A.~G., \& Kniazev, A.~Y.~2004, \aap, 425, 51P

\bibitem[Reines et al.~(2011)]{reines2011}
Reines, A.~E., Sivakoff, G~R., Johnson, K.~E., Brogan, C.~L.~2011, \nat, 470, 66

\bibitem[Reines et al.(2013)]{2013ApJ...775..116R} Reines, A.~E., Greene, 
J.~E., \& Geha, M.\ 2013, \apj, 775, 116

\bibitem[Satyapal et al.~(2007)]{satyapal2007}
Satyapal, S.,Vega, D., Heckman, T., O'Halloran, B., \& Dudik, R.~2007, \apj, 663, L9

\bibitem[Satyapal et al.~(2008)]{satyapal2008}
Satyapal, S., Vega, D., Dudik, R.~P., Abel, N.~P., \& Heckman, T.~2008, \apj, 677, 926

\bibitem[Satyapal et al.~(2009)]{satyapal2009}
Satyapal, S., et al.~2009, \apj, 704, 439

\bibitem[Schramm et al.~(2013)]{schramm2013}
Schramm, M., et al.~2013, \apj, 773, 7

\bibitem[Secrest et al.~(2012)]{secrest2012}
Secrest, N.~J., et al.~2012, ApJ, 753, 38

\bibitem[Secrest et al.~(2013)]{secrest2013}
Secrest, N.~J., et al.~2013, in preparation

\bibitem[Shields et al.~(2008)]{shields2008}
Shields, J.~C., et al.~2008, \apj, 682, 104

\bibitem[Simard et al.~(2011)]{simard2011}
Simard, L., Mendel, J.~T., Patton, D.~R., Ellison, S.~L., \& McConnachie, A.~W.~2011, \apjs, 196, 11

\bibitem[Simmons et al.~(2013)]{simmons2013}
Simmons, B., et al.~2013, \mnras, 429, 2199S

\bibitem[Stasinska et al.~(2006)]{stasinska2006}
Stasinska, G., Cid Fernandes, R., Mateus, A., Sodre, L., \& Asari, N.~V.~2006, \mnras, 371, 972

\bibitem[Stern et al.~(2012)]{stern2012}
Stern, D. et al.~2012, \apj, 753, 18

\bibitem[Teng et al.~(2005)]{teng2005}
Teng, S.~H., et al.~2005, \apj, 633, 664

\bibitem[Thuan \& Martin (1981)]{thuan81}
Thuan, T.~X., \& Martin, G.~E.~1981, \apj, 247, 823

\bibitem[Thuan et al.~(1999)]{Thuan1999}
Thuan, T.~X., Sauvage, M., \& Madden, S.~1999, \apj, 516, 783

\bibitem[van Wassenhove et al.~(2010)]{vanwassenhove2010}
van Wassenhove, S., Volonteri, M., Walker, M.~G., \& Gair, J.~R.~2010, \mnras, 408, 1139

\bibitem[Veilleux et al.~(2009)]{veilleux2009}
Veilleux, S., et al.~2009, \apjs, 182, 628V

\bibitem[Volonteri (2010)]{volonteri2010}
Volonteri, M.~2010, \aapr, 18, 27

\bibitem[Wright et al.~(2010)]{wright2010}
Wright et al.~2010, \aj,140,1868

\bibitem[Yan et al.~(2012)]{Yan2012}
Yan et al.~2012, \aj, in press


\end{thebibliography}
\end{document}